\documentclass[twocolumn]{aastex62}

\usepackage{xspace}
\newcommand{\src}{4U~1608$-$52\xspace}

\newcommand{\nicer}{\emph{NICER}}

\newcommand{\fluxcgs}{erg~s$^{-1}$~cm$^{-2}$}

\graphicspath{{./}{figures/}}

\received{XX, 2020}
\revised{XX, 2020}
\accepted{XX, 2020}

\submitjournal{ApJ}

\shorttitle{Thermonuclear X-ray Bursts with Late Secondary Peaks from 4U~1608$-$52}
\shortauthors{G\"uver et al.}

\begin{document}

\title{Thermonuclear X-ray Bursts with late secondary peaks observed from 4U~1608$-$52}

\correspondingauthor{Tolga G\"uver}
\email{tolga.guver@istanbul.edu.tr}

\author[0000-0002-3531-9842]{Tolga G\"uver}
\affiliation{Istanbul University, Science Faculty, Department of Astronomy and Space Sciences, Beyaz\i t, 34119, \.Istanbul, Turkey}
\affiliation{Istanbul University Observatory Research and Application Center, Istanbul University 34119, \.Istanbul Turkey}

\author[0000-0002-4729-1592]{Tu\u{g}ba Boztepe}
\affiliation{Istanbul University, Graduate School of Sciences, Department of Astronomy and Space Sciences, Beyaz\i t, 34119, \.Istanbul, Turkey}

\author{Ersin G\"o\u{g}\"u\c{s}}
\affiliation{Faculty of Engineering and Natural Sciences, Sabanc\i~University, Orhanl\i-Tuzla 34956, \.Istanbul, Turkey }

\author{Manoneeta Chakraborty}
\affiliation{Discipline of Astronomy, Astrophysics and Space Engineering (DAASE),
Indian Institute of Technology Indore, Khandwa Road, Simrol, Indore 453552, India}

\author[0000-0001-7681-5845]{Tod E. Strohmayer}
\affiliation{Astrophysics Science Division and Joint Space-Science Institute, NASA's Goddard Space Flight Center, Greenbelt, MD 20771, USA}

\author{Peter Bult}
\affiliation{Department of Astronomy, University of Maryland, College Park, MD 20742, USA} \affiliation{Astrophysics Science Division, NASA Goddard Space Flight Center, Greenbelt, MD 20771, USA}

\author[0000-0002-3422-0074]{Diego Altamirano} 
\affiliation{School of Physics and Astronomy, University of Southampton, Southampton, SO17 1BJ, UK}

\author[0000-0002-6789-2723]{Gaurava K. Jaisawal}
\affiliation{National Space Institute, Technical University of Denmark, 
   Elektrovej 327-328, DK-2800 Lyngby, Denmark}

\author[0000-0003-1188-4692]{Tu\u{g}\c{c}e Kocab\i y\i k}
\affiliation{Istanbul University, Graduate School of Sciences, Department of Astronomy and Space Sciences, Beyaz\i t, 34119, \.Istanbul, Turkey}

\author[0000-0002-0380-0041]{C.~Malacaria}\thanks{NASA Postdoctoral Fellow}
\affiliation{NASA Marshall Space Flight Center, NSSTC, 320 Sparkman Drive, Huntsville, AL 35805, USA}
\affiliation{Universities Space Research Association, Science and Technology Institute, 320 Sparkman Drive, Huntsville, AL 35805, USA}

\author{Unnati Kashyap}
\affiliation{Discipline of Astronomy, Astrophysics and Space Engineering (DAASE),
Indian Institute of Technology Indore, Khandwa Road, Simrol, Indore 453552, India}

\author{Keith C. Gendreau}
\affiliation{Astrophysics Science Division, NASA Goddard Space Flight Center, Greenbelt, MD 20771, USA}

\author{Zaven Arzoumanian}
\affiliation{Astrophysics Science Division, NASA Goddard Space Flight Center, Greenbelt, MD 20771, USA}
\author{Deepto Chakrabarty}
\affiliation{MIT Kavli Institute for Astrophysics and Space Research, Massachusetts Institute of Technology, Cambridge, MA 02139, USA}



\begin{abstract}

We report the temporal and spectral analysis of three thermonuclear X-ray bursts from \src, observed by the Neutron Star Interior Composition Explorer (NICER) during and just after the outburst observed from the source in 2020. In two of the X-ray bursts, we detect secondary peaks, 30 and 18 seconds after the initial peaks. The secondary peaks show a fast rise exponential decay-like shape resembling a thermonuclear X-ray burst. Time-resolved X-ray spectral analysis reveals that the peak flux, blackbody temperature, and apparent emitting radius values of the initial peaks are in agreement with X-ray bursts previously observed from \src, while the same values for the secondary peaks tend toward the lower end of the distribution of bursts observed from this source. The third X-ray burst, which happened during much lower accretion rates did not show any evidence for a deviation from an exponential decay and was significantly brighter than the previous bursts. We present the properties of the secondary peaks and discuss the events within the framework of short recurrence time bursts or bursts with secondary peaks. We find that the current observations do not fit in standard scenarios and challenge our understanding of flame spreading.
\end{abstract}



\section{Introduction} 
\label{sec:intro}

Neutron stars in low mass X-ray binaries (LMXBs) often exhibit sudden flashes of X-rays, called type-I X-ray bursts. These events, typically last only about tens of seconds and show a characteristic morphology, that is, fast rise and exponential decay. Time-resolved X-ray spectral studies of these events show that the burst spectra can be described with an evolving Planckian function (kT$\sim$1–3 keV) and the total energy released in such bursts can be anywhere between $10^{38-39}$~erg (see \cite{1993SSRv...62..223L}, \cite{2006csxs.book..113S} and \cite{2017arXiv171206227G} for detailed reviews on type-I X-ray bursts).

X-ray bursts are attributed to the unstable fusion of hydrogen and/or helium present in the material accreted onto the neutron star, therefore, they are also termed as thermonuclear X-ray bursts. \cite{Galloway2008} and \cite{2020ApJS..249...32G} provide up-to-date catalogs and analyses of an extensive sample of thermonuclear bursts spanning a wide range of characteristics. The burning primarily depends on the mass accretion rate to the surface \citep{1981ApJ...247..267F}, although various other physical factors like metallicity and the compactness of the neutron star also play a role \citep{1998ASIC..515..419B, 2003ApJ...599..419N}. 

Almost all of the existing scenarios suggest that the material needs to accumulate for a certain period of time for the next X-ray burst to happen and most of the fuel is burned during a burst, with only a thin layer of ashes remaining \citep{1987ApJ...319..893L, 1988MNRAS.233..437V, 2008ApJS..174..261F, 2010ApJS..189..204J}. The recurrence time of X-ray bursts depend strongly on the mass accretion rate and typically can be hours or longer. However,  X-ray bursts with even shorter recurrence times have been observed. These events are generally referred to as short waiting time (SWT) bursts, where the waiting time is defined to be less than 45 minutes \citep{2017ApJ...842..113K} however, bursts with recurrence times as short as only a few minutes have been observed from several sources including EXO~0748$-$676 \citep{1987ApJ...323..575G,2007A&A...465..559B,2009A&A...497..469I}, GS~0836$-$429 \citep{1992PASJ...44..641A}, \src \citep{1980ApJ...240L.143M}, 4U 1705$-$440 \citep{1987ApJ...323..288L}, the Rapid Burster \citep{2014MNRAS.437.2790B}, GRS 1747$-$312 \citep{2003A&A...409..659I} and several others \citep{2010ApJ...718..292K,2012ApJ...748...82L}.  In some cases triple or even quadruple SWT bursts have been observed
\citep{2010ApJ...718..292K} including an event observed from \src \citep{2009MNRAS.398..368Z}.

The SWTs between  X-ray bursts are not very well understood within the framework of classical thermonuclear flash scenarios, because of the fact that the waiting times in between the  X-ray bursts are often too short to accrete enough material and start another ignition. The secondary bursts within these events are thought to be caused by the burning of residual fuel that remained from the primary burst. This was first demonstrated by \citep{1981ApJ...247..267F} using several mechanisms like an incomplete nuclear flame, fuel storage, and a mixing mechanism with new accreted material. \cite{2017ApJ...842..113K} suggested that second bursts must be powered by fuel remaining from the previous explosion given the fact that these events are typically less bright and have shorter durations. SWT burst events have been observed more frequently in sources where the accreted material is hydrogen-rich and the neutron star spin frequency is larger than $>$500 Hz \citep{2010ApJ...718..292K}. \cite{2007A&A...465..559B} suggested that stochastic processes associated with fast rotation may play an important role in the occurrence of SWT burst events. 

In addition to bursts with short recurrence times, there is also another group of X-ray bursts where secondary peaks are observed, these events are called double-peaked bursts \citep[see, e.g.][]{1985ApJ...299..487S,2006ApJ...641L..53B, 2006ApJ...636L.121B}. Some examples of observations of these rare events include  X-ray bursts from 4U~1636$-$536 \citep{2009MNRAS.398..368Z}, \src \citep{1989A&A...208..146P,2017arXiv171206227G,Jaisawal2019}, GX~17$+$2 \citep{2002A&A...382..947K} and 4U~1709$-$267 \citep{2004MNRAS.354..666J}.  Unlike in standard type-I bursts, during these events, X-ray intensity reaches to a peak followed by a decline and another subsequent rise. The secondary peak can reach to similar intensity levels as the initial peak and the peaks are separated by a few seconds. Most recently \cite{2020MNRAS.tmp.3345L} studied 16 multi-peaked X-ray bursts observed from 4U~1636$-$536 with Rossi X-ray Timing Explorer (RXTE). They find an anti-correlation between the peak flux of the secondary peaks and the separation time between the peaks. They also find that the ratio of the peak fluxes of the peaks in bursts are correlated with the temperature of the thermal component in the pre-burst spectra and conclude that double peaks maybe related to the accretion rate in the disc or the temperature of the neutron star.

\citet{2006ApJ...641L..53B, 2006ApJ...636L.121B} suggested that in these cases, the X-ray burst is ignited at or around one of the poles and propagates towards the equator. As the burning front propagates towards the equator, either due to acting Coriolis forces \citep{2007ApJ...666L..85B} or due to the effect of the magnetic field at the surface \citep{2006ApJ...641..471P} it stalls for a few seconds, causing a decline in the X-ray intensity. 
As the burning front continues to propagate towards the opposite pole the intensity increases again. 
Note that, \citet{2013MNRAS.434.3526C, 2015MNRAS.448..445C} investigate possible mechanisms for stalling near the equator and conclude that, the Coriolis effects alone may not be enough to stall the front, however, their simulations do not take into account the effects of ongoing accretion and magnetic fields. 
As an alternative, \cite{2004ApJ...608L..61F} suggest that the stall may be caused by the waiting points in the nuclear reaction chain, however, such a scenario falls short of a complete explanation of the burst profile characteristics.

The transient low mass X-ray binary \src is a well-known X-ray burster, since its first detection with the two Vela-5 satellites \citep{1976ApJ...206L.135B}.  \src is classified as an atoll source based on spectral and timing properties by \cite{1989A&A...225...79H}. The detection of the burst oscillations from the source suggests that the spin period of the neutron star is $\sim$620 Hz \citep{2002ApJ...580.1048M}. \cite{2002ApJ...568..901W} used the observed periodic modulation in the $I$ band data to claim an orbital period for the system as 12.9~h. \cite{2002ApJ...568..901W} argued that if the companion is assumed to be a main-sequence star, then existing observations indicate an F to G type donor. Using the observed X-ray bursts with RXTE, first \cite{2010ApJ...712..964G} and later \cite{Ozel2016} determined a number of physical parameters of the system. Using red clump giants and soft X-ray observations, the distance of \src is determined to be most likely at 4~kpc or even larger \citep{Ozel2016}. The Eddington limit of the source was also measured at 3.54$\pm$0.38$\times$10$^{38}$~erg~s$^{-1}$ (or 18.5$\pm$2.0$\times$10$^{-8}$~\fluxcgs) using the touchdown fluxes measured during thermonuclear X-ray bursts. Throughout the paper, we use these values as our reference.

The Multi-INstrument Burst ARchive (MINBAR) reports 147 bursts in the RXTE, BeppoSAX, and INTEGRAL archives from this source \citep{2020ApJS..249...32G}. Using Hakucho, \citet{1980ApJ...240L.143M} detected X-ray bursts separated by about 10 minutes. Furthermore, \citet{2010ApJ...718..292K} reported the detection of SWT burst events, using an earlier version of the MINBAR catalog. In the final version of the MINBAR catalog \citep{2020ApJS..249...32G} there are several bursts separated by as short as~6 minutes (with MINBAR burst IDs: 2218, 1619, 2885, 7497, 7498) observed either by RXTE/PCA or INTEGRAL JEM-X. \citet{Jaisawal2019} reported the detection with NICER of a secondary peak in the cooling tail of an X-ray burst showing photospheric radius expansion, in a soft state. The soft X-ray sensitivity of NICER allows for the comparison and testing of models based on the effects of absorption to those that rely on stalling of flame propagation or additional burning \citep{Jaisawal2019}.

Starting from May 27th to roughly 19th of August 2020, \src has been in an outburst state. This outburst has been monitored by Swift/BAT and the Monitor of All-sky X-ray Image \citep[MAXI,][]{2009PASJ...61..999M}.
Starting from June 4th, NICER performed a high cadence campaign to monitor the evolution of \src during the whole outburst. A detailed analysis of the NICER dataset regarding the spectroscopic evolution of the system during the outburst is ongoing and will be reported in a forthcoming paper. Within these observations we detected two X-ray bursts each showing clear secondary peaks, ~30 and ~18 seconds after the initial peaks and another third burst just after the outburst. In this paper we concentrate on the temporal and spectral analyses of these X-ray bursts, especially the secondary peaks and their nature. 


\section{Observation and Data Analysis} \label{sec:analysis}

\begin{figure*}
    \centering
\includegraphics[scale=0.4]{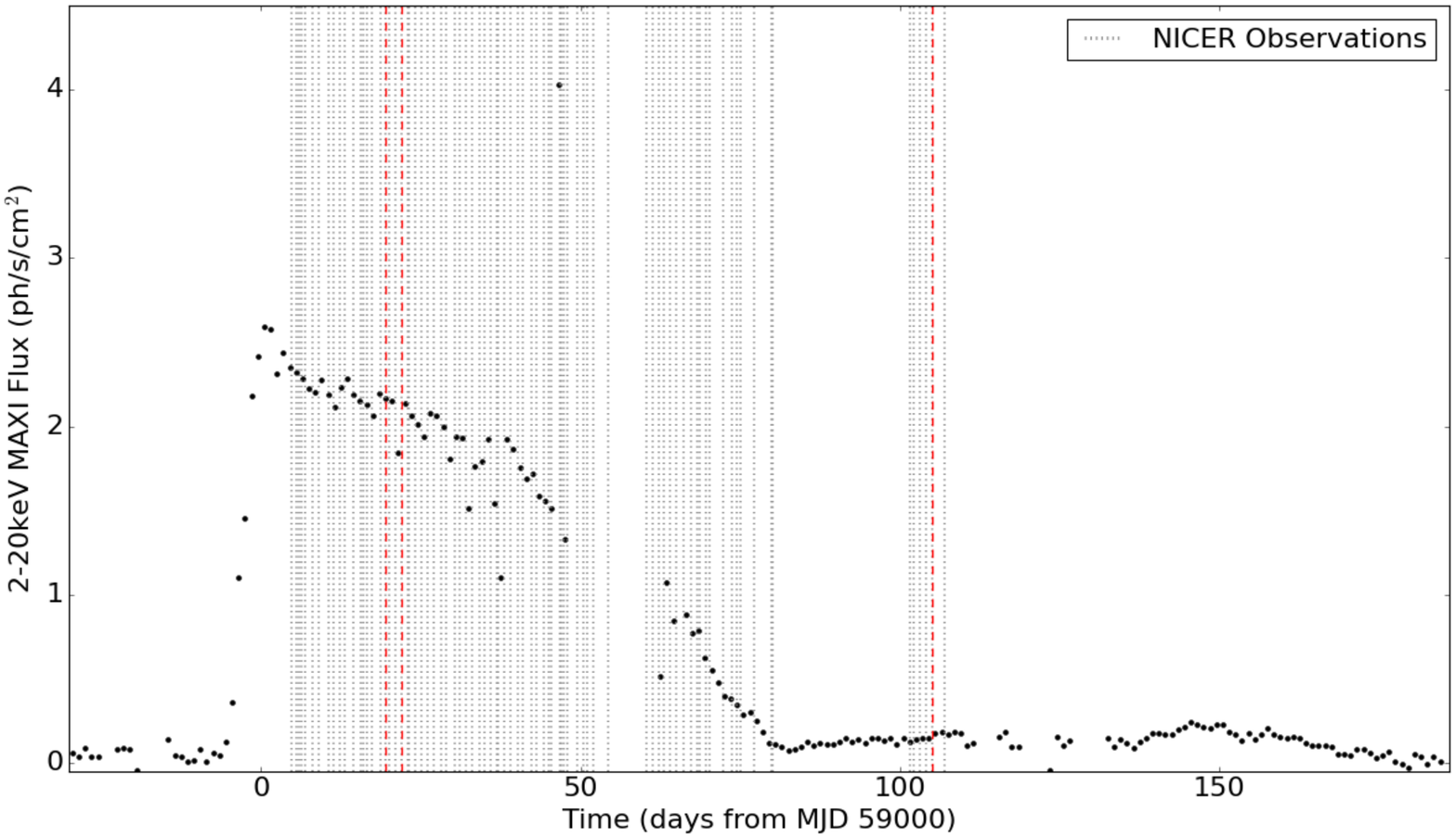}
    \caption{2$-$20~keV MAXI lightcurve of \src from daily average measurements, starting from May (05/01/2020) to December (12/03/2020). The grey dotted lines indicate the dates of \nicer observations and the red dashed lines shows \nicer~burst times. The sudden rise seen in the MAXI lightcurve around 50 days after the start of the outburst is also shown and is a superburst not reported before (Iwakiri et al. 2021 in preparation).}
    \label{fig:maxi_lc}
\end{figure*}

On-board the International Space Station (ISS), NICER contains 56 (52 of which were active during the observations presented here) X-ray concentrator optics and silicon drift detector pairs to provide a large effective area in the 0.2$-$12~keV band \citep{2012SPIE.8443E..13G, 2017AAS...22930903G}. According to MAXI, \src started an outburst on May 27th, 2020. NICER started observing the source on June 4th, 2020 and continued to monitor the evolution of the outburst comprehensively. MAXI lightcurve and the NICER coverage is shown in Figure \ref{fig:maxi_lc} together with the detected thermonuclear bursts. A total unfiltered exposure time of~283~ks, with 78 different observations is obtained with NICER, up to the 14 September 2020 (see Figure \ref{fig:maxi_lc}). These observations cover the ObsIDs, 3050070101, and all observations with ids 365702xxxx, where xxxx changes from 0101 to 9906.

After applying standard filtering criteria, using HEASOFT v.6.27.2, NICERDAS v7a, and the calibration files as of 2020/07/27, we had a total of 174~ks clean exposure time. We searched for X-ray bursts including the unfiltered data and found that within this rich data set, NICER detected three type-I X-ray bursts that occurred on June 23rd and 26th at around the peak of the outburst and on September 13th just after the end of the outburst. These bursts were detected in the observations with ObsIDs 3657021501, 3657021801, and 3657029905, which had 2.6~ks, 5.3~ks, and 232 s of effective exposure times, respectively. Note that the standard screening of the data eliminates the time of the last  X-ray burst. We, therefore, used the unfiltered data for the analysis of this burst. Throughout the paper, we refer to these X-ray burst events as bursts I, II and III and call the two peaks in the first two events as the initial and secondary. We present the lightcurves of these X-ray bursts in Figure \ref{fig:burst_lc} and further discuss below.

\begin{figure}
    \centering
     \includegraphics[scale=0.5]{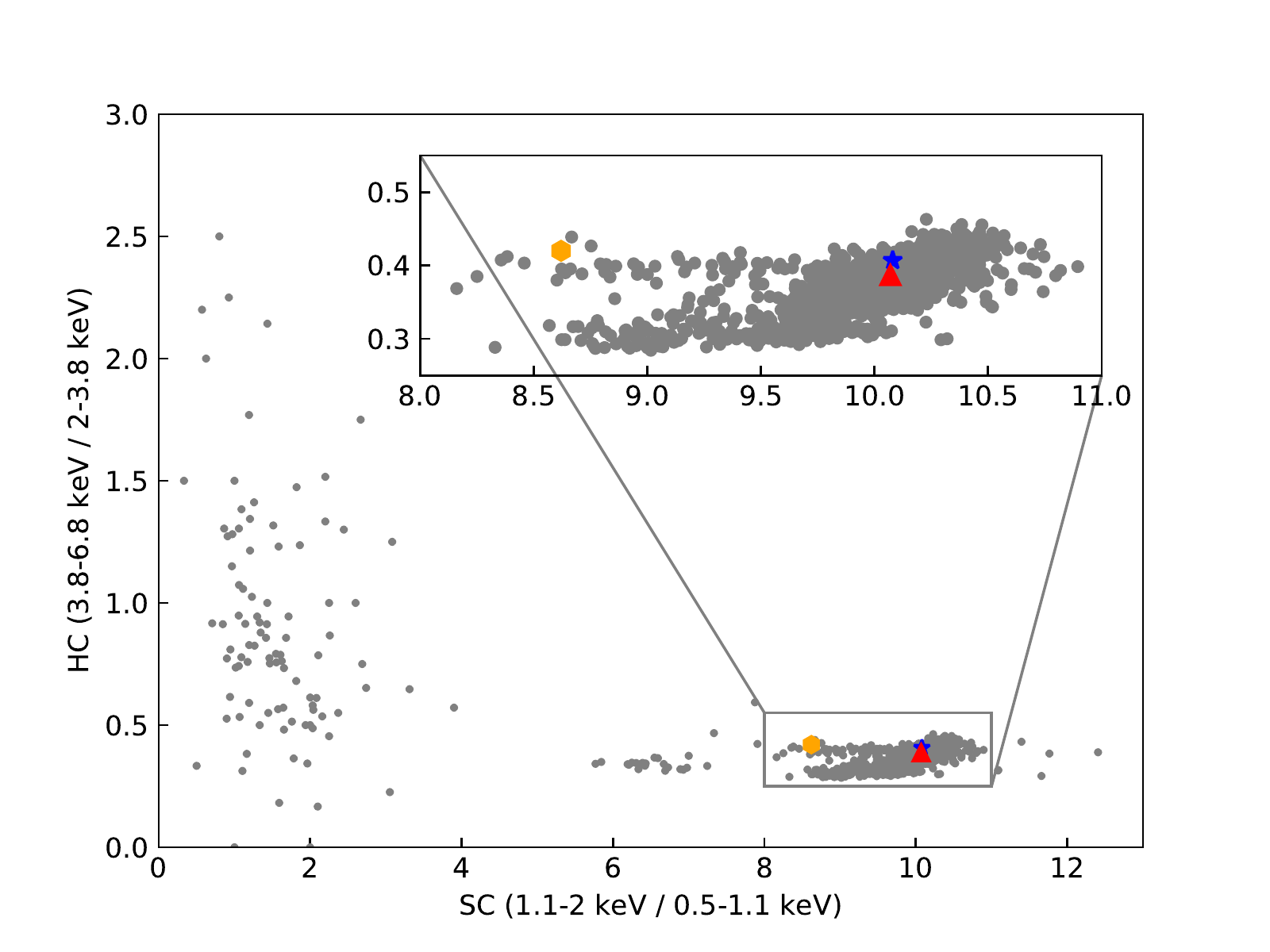}
    \includegraphics[scale=0.5]{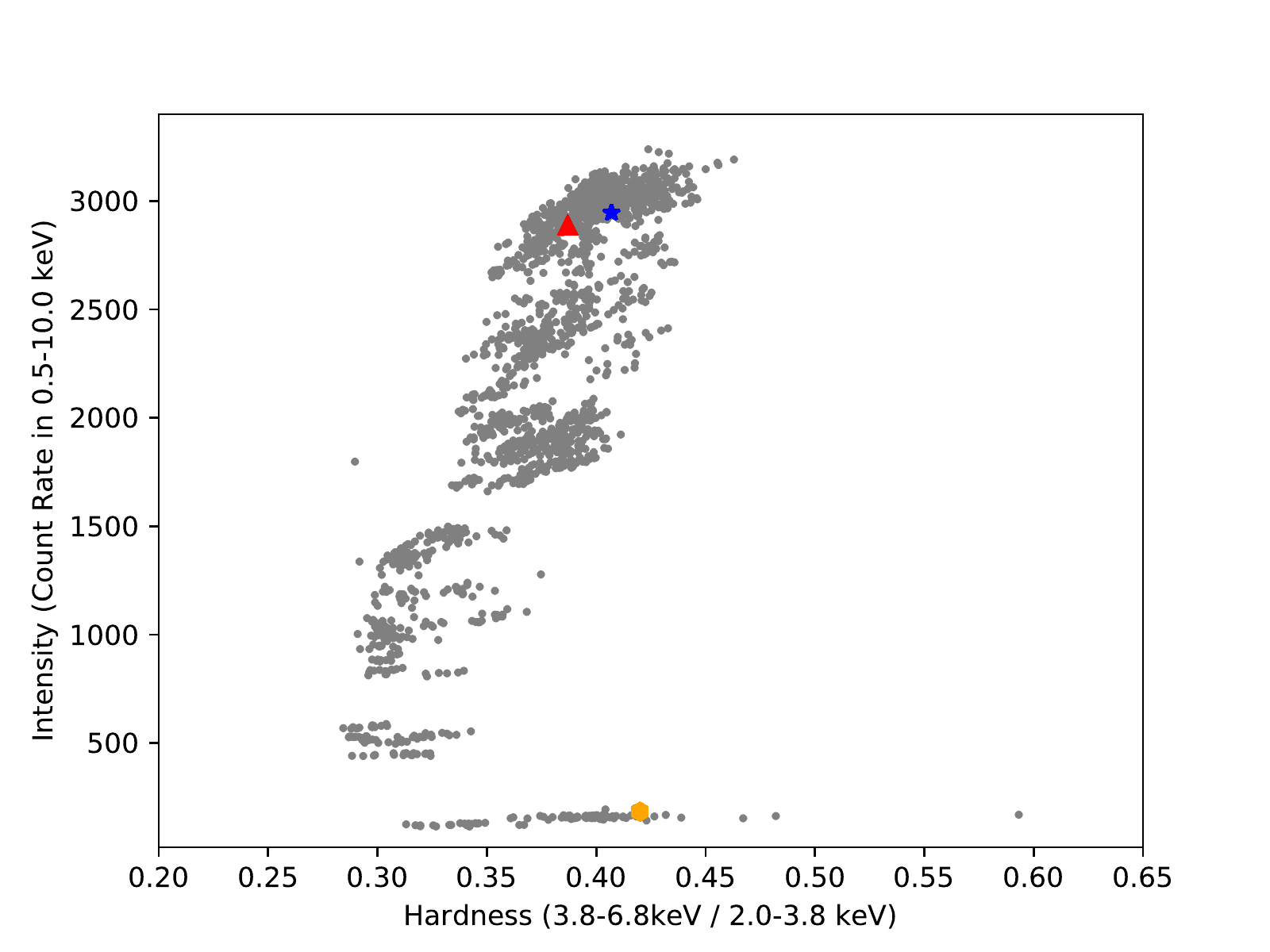}
    \caption{Hardness-Intensity diagram extracted from all of the observations in which we searched for an X-ray burst during the outburst. Detected X-ray bursts are shown with blue star, red triangle, and orange circle symbol for burst I, II, and III, respectively.}
    \label{fig:hid_ccd}
\end{figure}

An in-depth analysis of the outburst will be presented elsewhere, however to put the observed X-ray bursts in perspective we show in Figure \ref{fig:maxi_lc} the $2-20$~keV lightcurve of the outburst as observed by MAXI starting from 1st of May to November 11th 2020. Compared to the outbursts observed recently, the 2020 outburst has been one of the longest and brightest events on record. Using X-ray lightcurves extracted from the analyzed \nicer~ data binned to have a time resolution of 128~s, we also calculated X-ray hardness-intensity and color-color diagrams following the energy selection of \cite{Jaisawal2019}. We present these diagrams in Figure \ref{fig:hid_ccd}.

\subsection{X-ray lightcurves}

\begin{figure*}
    \centering
    \includegraphics[scale=0.36]{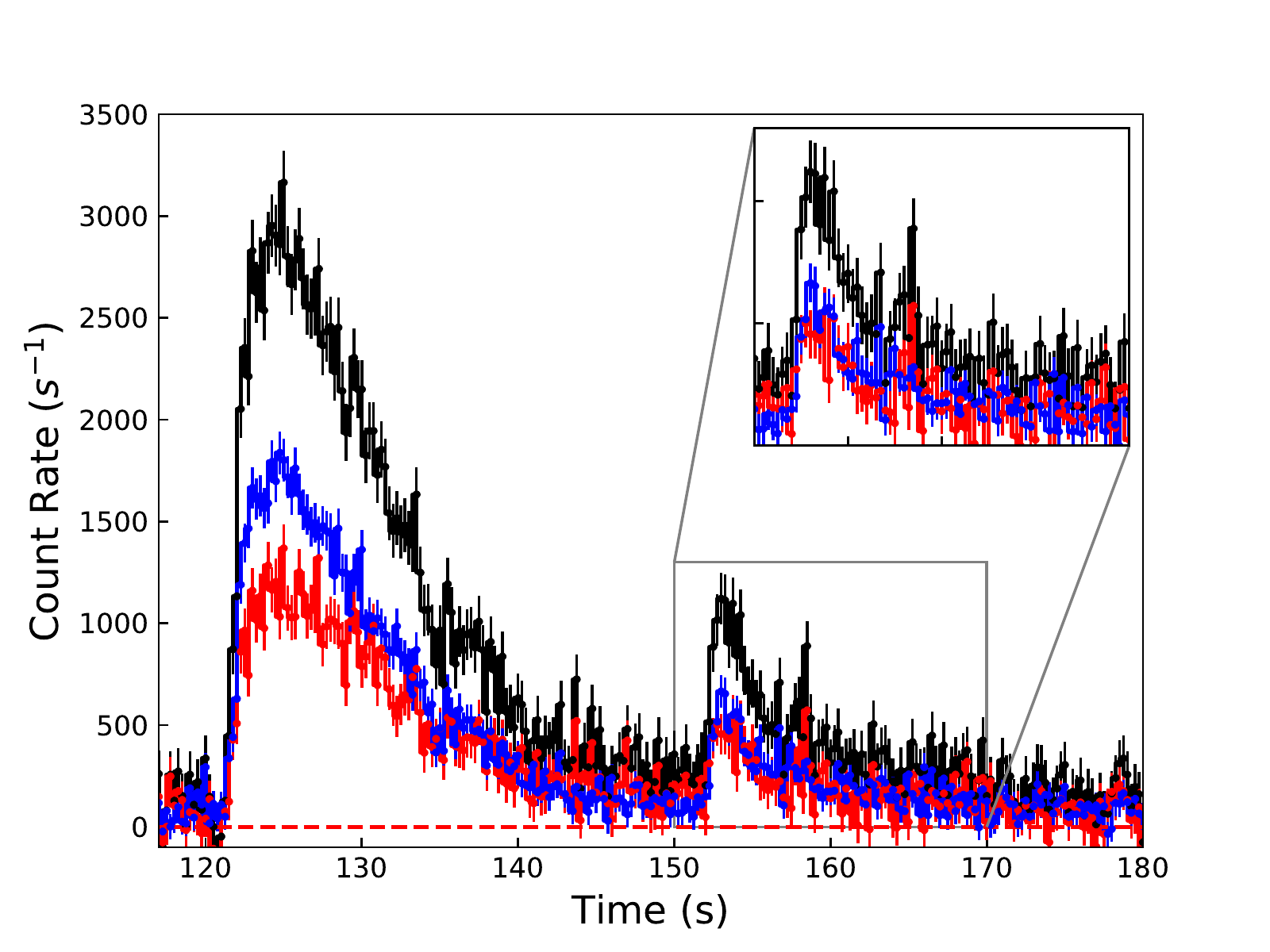}
    \includegraphics[scale=0.36]{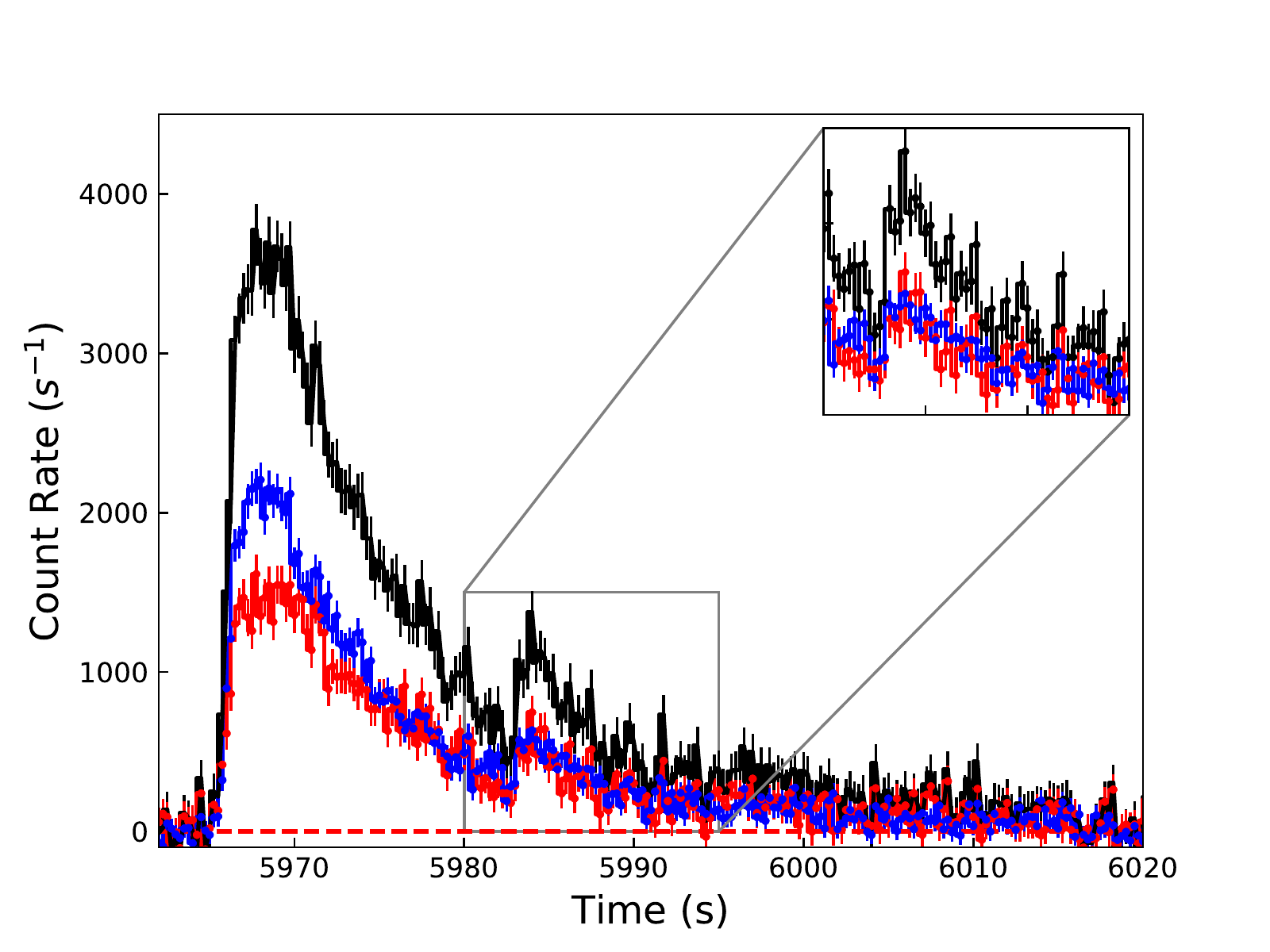}
        \includegraphics[scale=0.36]{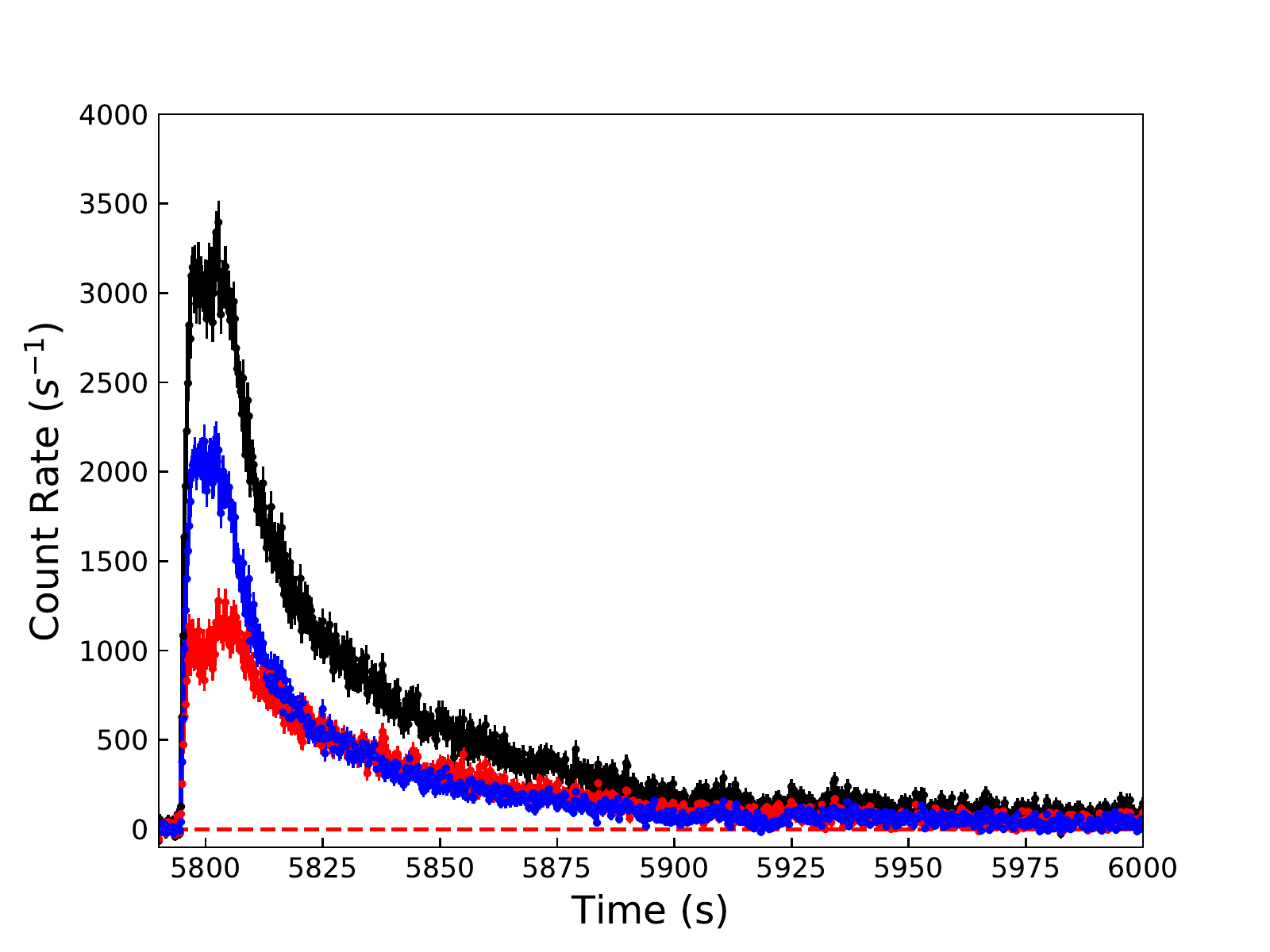}
    \caption{The X-ray lightcurve of burst~1, burst~2 and burst~3 from \src. 0.3$-$12, 3.0$-$12.0, and 0.3$-$3.0~keV lightcurves are shown in black, blue and red, respectively. The persistent count-rate is subtracted from each ligthcurve.}
    \label{fig:burst_lc}
\end{figure*}

We applied barycenter corrections to the event lists using the \emph{barycorr} tool, JPL DE430 planetary ephemeris and the coordinates of the source as R.A. 16$^h$12$^m$43$^s$, decl. 52$^\circ$ 25$'$ 23.$\arcsec$2 (J2000). Using the barycentred events we generated 0.3$-$3.0, 3.0$-$12.0, 0.3$-$12.0,~keV lightcurves, with a time resolution of 0.25~s (see Figure \ref{fig:burst_lc}). The X-ray  bursts are clearly visible 120, 5964, and 5795 seconds after the start of the observations, respectively. In the first two burst, a secondary peak is also clearly visible in all bands, while the last burst lack any evidence of such a structure. In burst I, a secondary peak or possibly a second burst is visible roughly 30~s after the initial burst started. In burst II, the secondary peak is even closer to the initial peak and starts $\sim 18$~s after the burst.

We modeled the $0.3-12$ keV burst lightcurves of \src to better characterize the observed bursts. We determined the pre-burst count rates calculating the average of data 50~s before each burst. For the secondary peak in burst I, we determined the average using the data 8~s data prior to it. These values, together with the peak count rates are presented in Table \ref{burst_charac} and show that the initial peak did not yet fully return to the persistent count rate, but very close to it. In burst II the secondary peak happens during the tail. In each case the peak count rates of the secondary peaks reached two-thirds of the initial peaks.

We calculated the rise times for each X-ray burst defining them as the time between the first bin within 1$\sigma$ of the peak count-rate and the last time bin within 1$\sigma$ of the persistent rate. In order to determine the exponential decay time-scales, we fit the decay of the X-ray bursts with an exponential function plus a constant, with the constant fixed to the pre-burst level. Note that to better model the decay of burst II, we removed a 10-second interval covering the secondary peak. The total decay of burst III lasted for ~205~s if we measure from peak to the pre-burst level. This whole interval can be modelled with a double exponential but to be able to better compare with the previous X-ray bursts we only modeled the decay of the first 50~s after the peak.  The best fit values are given in Table \ref{burst_charac} and further underline the fast rise and exponential decay nature of the secondary peaks themselves.

\begin{table*}
\centering
\caption{Some characteristic properties of the bursts.}
\begin{tabular}{ccc|cc|c}
\hline
\hline
                &  \multicolumn{2}{c|}{ Burst I} & \multicolumn{2}{c|}{Burst II}  & Burst III \\
                &   Initial Peak    & Secondary Peak & Initial Peak & Secondary Peak &  \\
                \hline
Time (MJD)      & \multicolumn{2}{c|}{59019.576478}  & \multicolumn{2}{c|}{59022.096125} & 59105.212414\\
Time Between Peaks (s)  &  \multicolumn{2}{c|}{28.00$\pm$0.25} &  \multicolumn{2}{c|}{18.00$\pm$0.25}&  -- \\
Prior Count Rate & 2978$\pm$8 & 3269$\pm$21 & 2893$\pm$8 &--& 63.7$\pm$1.0 \\
Persistent Flux$^a$ & \multicolumn{2}{c|}{2.99$\pm$0.01} & \multicolumn{2}{c|}{2.58$\pm$0.01} & 0.22$\pm$0.01 \\
Peak Count Rate$^b$ ($s^{-1}$) & 6096$\pm$156 & 4052$\pm$127& 6660$\pm$163 & 4264$\pm$131& 2240$\pm$94\\ 
Peak Flux$^a$ & 6.2$\pm$0.5 & 1.87$\pm$0.2 & 8.72$_{-0.8}^{+0.6}$ &1.4$\pm$0.2& 12.64$\pm$1.42 \\
Rise Time (s) & 1.4 & 0.76 &1.93& 1.75 &4.0 \\
$\tau$ &9.13$\pm$0.15 &10.3$\pm$0.5&9.6$\pm$0.3& 7.2$\pm$0.6&11.42$\pm$0.26\\ 
\hline
\end{tabular}
\label{burst_charac}\\ 
\footnotesize{$^a$  Unabsorbed bolometric flux in units of $\times$10$^{-8}$~\fluxcgs.}\\
\footnotesize{$^b$Including the persistent count rate of the source.}
\end{table*}


\subsection{Time Resolved Spectral Analysis of the Bursts} \label{sec:temporalanalysis}

To analyze the spectral evolution of the X-ray  bursts and better characterize the nature of the secondary peaks, we performed a time-resolved spectral analysis. For the extraction of the X-ray burst spectra we followed a method that is very similar to what has been done by \cite{Galloway2008} or \cite{2012ApJ...747...76G}. We extracted spectra with 0.5~s time resolution up to the peak of each X-ray burst. From the peak, we extracted X-ray spectra with exposure times of 0.5~s, 1.0~s or 2.0~s depending on the total count rate. We used the up to date response and ancillary response files as in NICER CALDB release \texttt{xti20200722}, however, note that we removed the data from Focal Plane Modules 14 and 34 and therefore adjusted the response and ancillary response files accordingly. 

Time-resolved spectral analysis of X-ray bursts is affected by the persistent emission of the low mass X-ray binary and needs to be taken into account carefully \citep{Galloway2008, 2012ApJ...747...76G, 2012ApJ...747...77G, 2013ApJ...772...94W, 2015ApJ...801...60W}. We extracted X-ray spectra prior to each X-ray burst with an exposure time of 100~s and modeled this pre-burst emission. For the spectral analysis, we utilized \emph{Sherpa}  \citep{2001SPIE.4477...76F} distributed with the CIAO v4.12. We created background files for each observation using the \texttt{nibackgen3C50}\footnote{\url{https://heasarc.gsfc.nasa.gov/docs/nicer/tools/nicer\_bkg\_est\_tools.html}} tool. For burst I and burst II, we fit the energy spectrum obtained from this interval with an absorbed disk blackbody (\emph{diskbb}) and a blackbody (\emph{bbodyrad}) model \citep{2003A&A...410..217G}. For burst III, however adding a blackbody to the disk blackbody component resulted in unconstrained parameters, we therefore used a power-law model, instead of a blackbody to better characterize the X-ray spectrum. Note that this last burst is observed at a low-hard state compared to the earlier bursts (see Figure \ref{fig:hid_ccd}). To take into account the Hydrogen column density we assumed the abundance of the interstellar medium and used the \emph{tbabs} model \citep{2000ApJ...542..914W}. With these models, we are able to obtain very good fits to the X-ray spectra in the $0.5-10$~keV range. We present the results of these fits in Table \ref{t_sp_res}.

\begin{table*}
\caption{Best fit model results for pre-burst X-ray spectra of \src.}
\begin{tabular}{ccccccccc}
\hline 
& N$_H$ & kT$_{in}$ & Inner Disk Radius & kT$_{BB}$ / $\Gamma$ & R$_{BB}$ & Flux* & $\chi^2_{\nu}$ / dof \\
& $\times$10$^{22}$cm$^{-2}$&keV&$km^2/D_{10kpc}^2$&keV&${km}^2/D_{10kpc}^2$& & \\
\hline
Pre Burst I&1.40$\pm$0.01&1.32$\pm$0.09&256.61$\pm$53.31&1.96$\pm$0.13&76.22$\pm$26.07&2.37$_{-0.08}^{+0.10}$&1.08/590&\\
Pre Burst II&1.40$\pm$0.01&1.17$\pm$0.07&372.19$\pm$67.06&1.73$\pm$0.08&111.91$\pm$29.83&2.28$\pm$0.01&1.11/666&\\ 
Pre Burst III&1.49$\pm$0.07&1.50$\pm$0.27&11.04$\pm$3.6&0.55$\pm$0.62&--&0.17$\pm$0.01&1.03/122\\ 
\hline
\end{tabular}
\label{t_sp_res}\\
\footnotesize{$*$ Unabsorbed 0.5$-$10~keV flux in units of $\times10^{-8}$~\fluxcgs.}\\
\end{table*}

We calculated the X-ray flux in the 0.01$-$200~keV band based on the best fit parameters and found the bolometric unabsorbed fluxes of the source prior to burst I, II, and III as F$_{bol}$=2.99$\pm$0.04$\times10^{-8}$ \fluxcgs, F$_{bol}$=2.58$\pm$0.01$\times10^{-8}$ \fluxcgs, and F$_{bol}$=0.22$\pm$0.01$\times10^{-8}$ \fluxcgs~ respectively. Using the Eddington limit from \cite{Ozel2016} these fluxes indicate that the persistent emission was 16.16\% and 13.95\%, 1.19\% Eddington. 

\begin{figure*}
    \centering
    \includegraphics[scale=0.5]{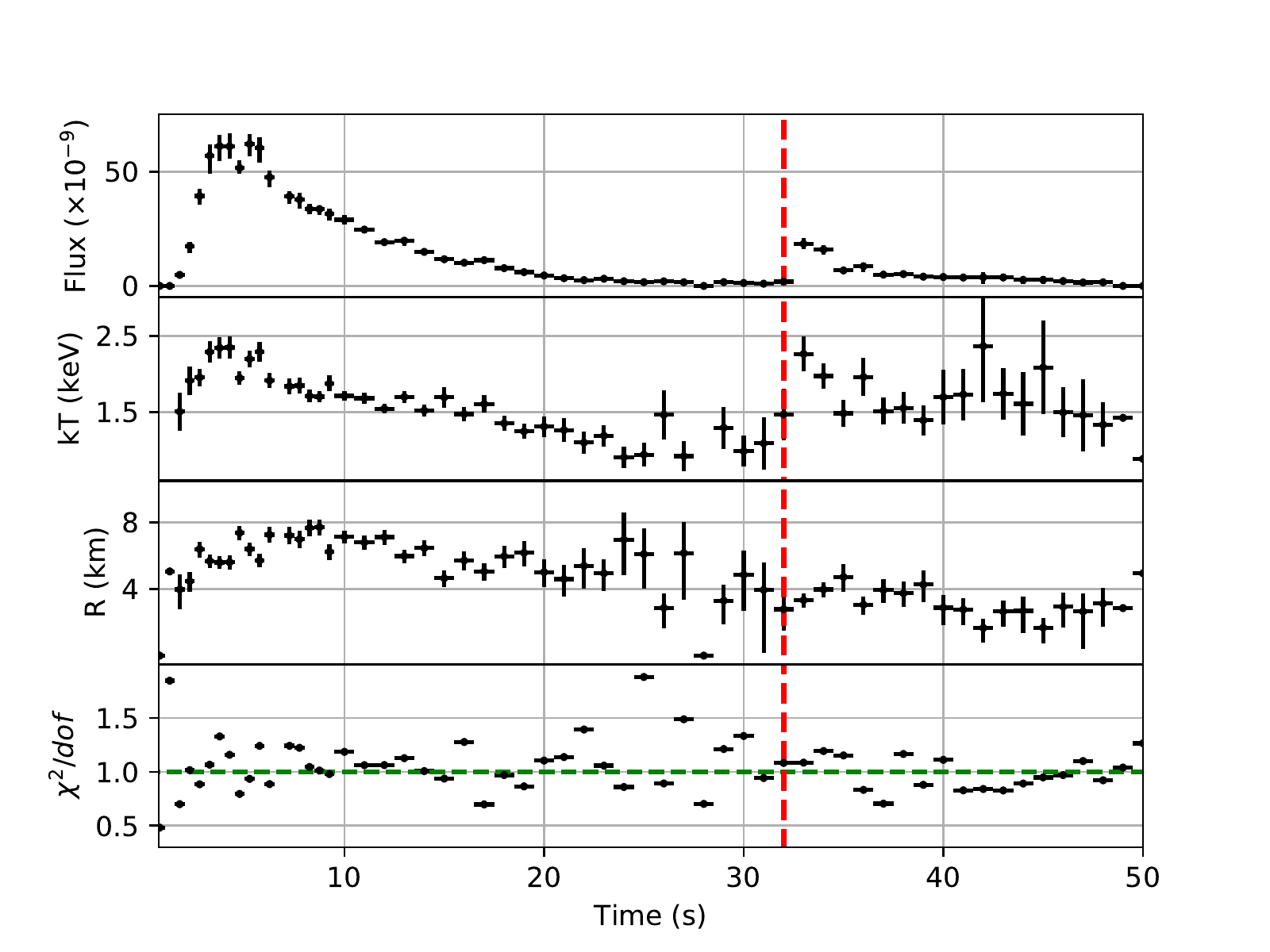}
    \includegraphics[scale=0.5]{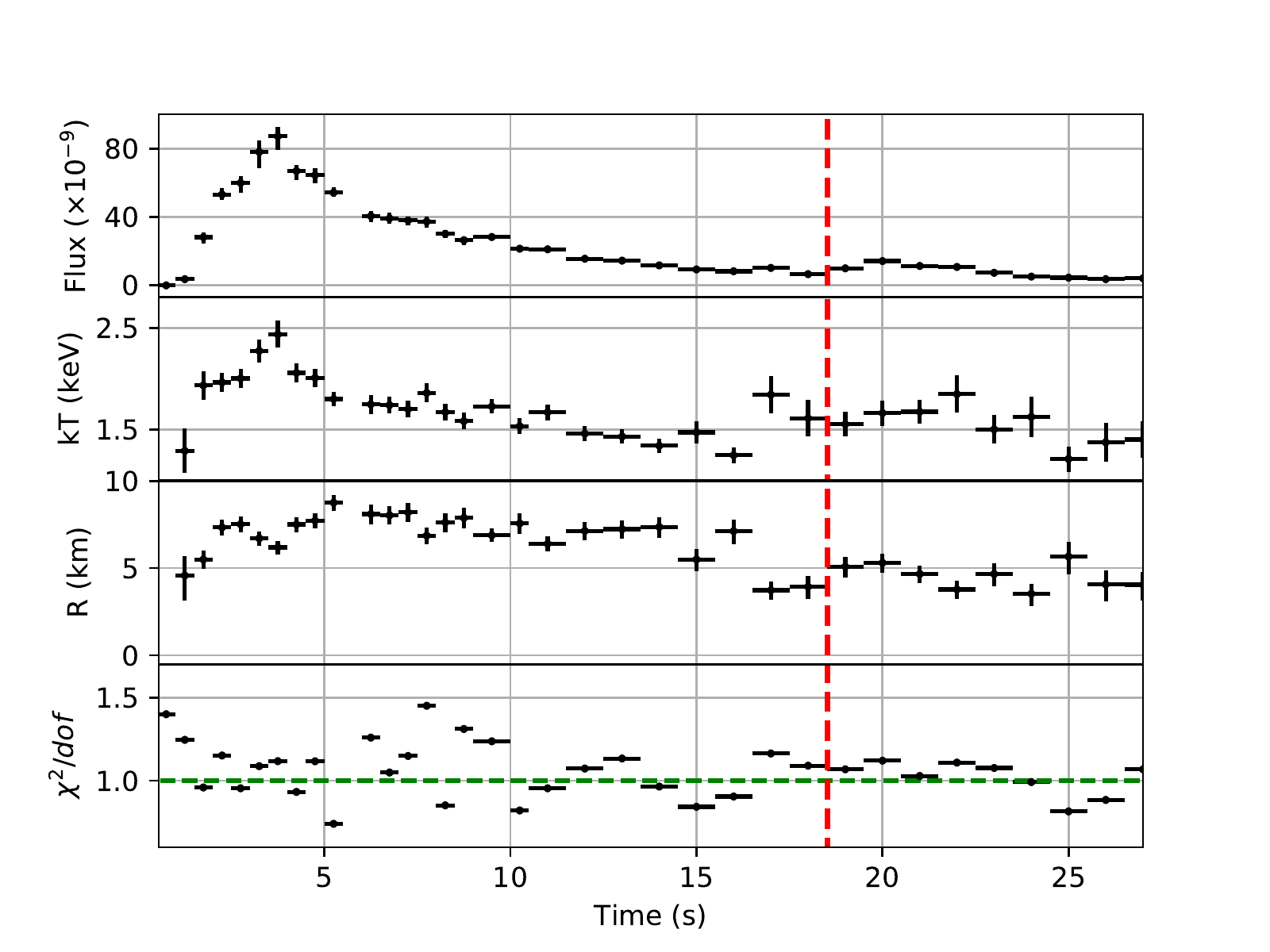}
    \includegraphics[scale=0.5]{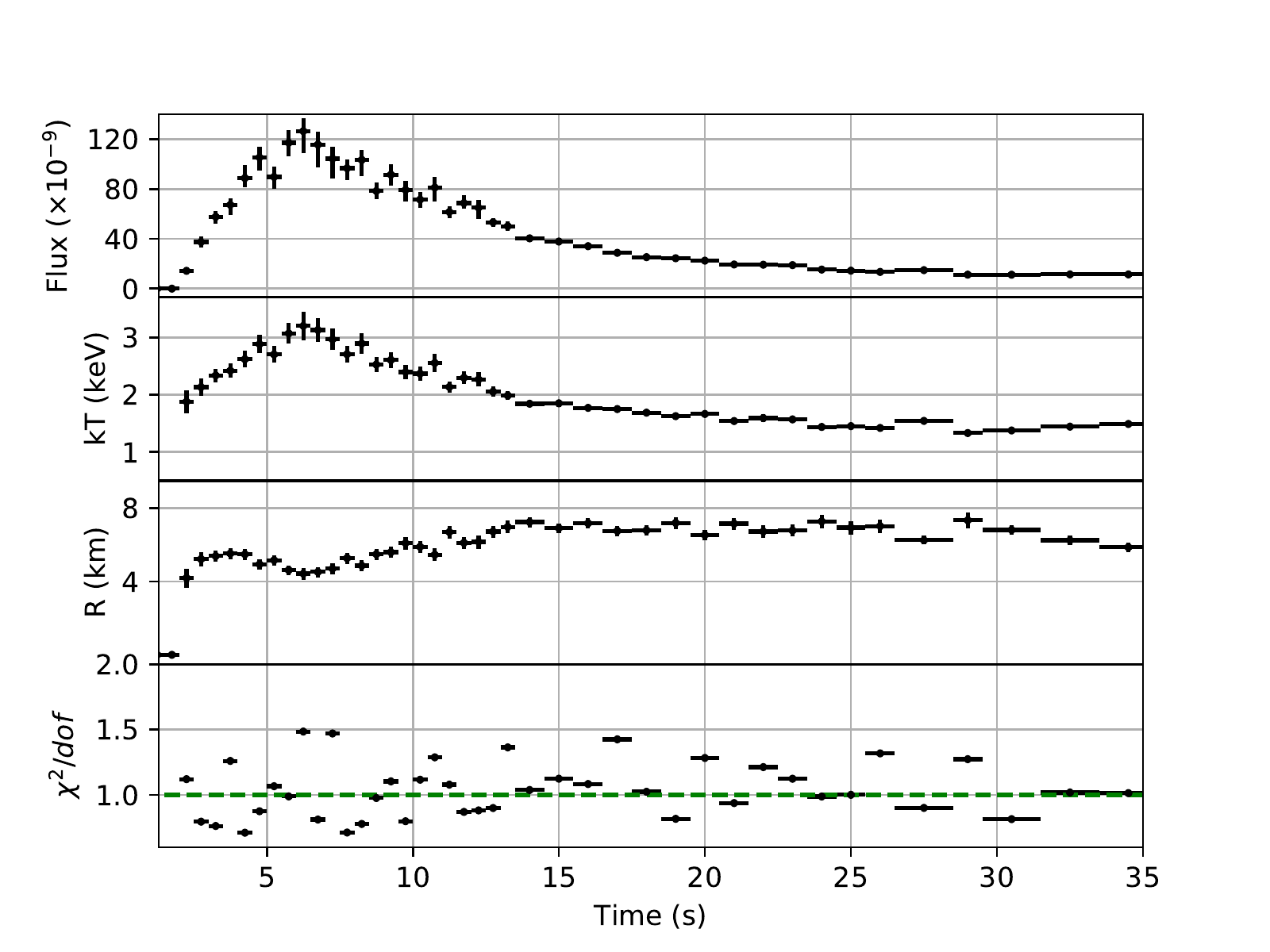}
    \caption{ Spectral evolution of burst I (left panel), burst II (right panel) and burst III based on blackbody model fits. Apparent emitting radius values are calculated assuming a distance of 4~kpc. Red vertical lines shows the start of the secondary peaks detected. In all panels fluxes are bolometric and given in units of \fluxcgs.}
    \label{fig:bursts_method1}

\end{figure*}

For the time-resolved spectroscopy of the X-ray burst emission, we used custom python scripts (using Astropy  \citep{2018AJ....156..123A}, NumPy \citep{van2011numpy}, Matplotlib \citep{Hunter:2007}, and Pandas \citep{reback2020pandas}) within \emph{Sherpa}. We grouped each spectrum to have at least 50 counts per channel and fit the 0.5$-$10.0 keV range. We used the best fit model for the persistent emission and subtracted only the instrumental background. On top of that, we added a blackbody component to account for the additional emission arising from the X-ray burst. Our fits generally resulted in fairly acceptable $\chi^{2}$ / dof values, with mean values of 1.06/59, 1.05/58, and 1.02/44 for the first 40~s of each burst. We also tried to add a second blackbody component \citep[as in][]{2019ApJ...885L...1B} or the so-called f$_{a}$ model \citep{2013ApJ...772...94W,2015ApJ...801...60W}, but apart from a few spectra these did not result in a significant improvement in the fits and resulted in unconstrained parameters for these additional models. We, therefore, present only the results for the blackbody models. Spectral evolution is shown in Figure \ref{fig:bursts_method1} for both bursts.

Time-resolved spectroscopy indicates that the observed X-ray bursts do not show any evidence for photospheric radius expansion. We calculated the bolometric fluxes by extrapolating the best-fit blackbody model to cover the 0.01 $-$ 200 keV range. We found that the peak bolometric fluxes reach to 6.2$\pm$0.5$\times10^{-8}$,  8.72$_{-0.8}^{+0.6}\times10^{-8}$, and 12.64$\pm$1.42$\times10^{-8}$~\fluxcgs, in each burst respectively. These values show that X-ray bursts reach roughly 33, 47, and 68\% of the Eddington limit determined based on earlier observations \citep{Ozel2016}. 
Our spectral analysis reveals a significant spectral variation associated with the secondary peaks at the moments we expect from the X-ray lightcurves, especially in the burst I. In burst II, the spectral variation is not as evident. Considering that this may be due to our exposure time determination, we also tried to arbitrarily shorten the exposure times of the X-ray spectra to 0.5~s to better resolve the spectral variations, however, this did not necessarily improve spectral results. This finding can also be due to the fact that the flux contribution exclusively due to this peak is significantly less than that of the secondary peak in burst I, resulting in a smaller impact on the spectral evolution. Any change due to the secondary peak may also be harder to accurately measure as it will be embedded in the spectral parameter evolution of the initial peak.

At the peak of burst I, the blackbody temperature and apparent emitting radius values reach kT$_{BB}$=2.19$\pm$0.11~keV and R=6.4$\pm$0.4~km (assuming a distance of 4~kpc). The secondary peak reaches a maximum flux of F$_{peak}$=1.87$\pm$0.2$\times$10$^{-8}$~\fluxcgs. We find that at the secondary peak, while the temperature reaches 2.26$\pm$0.22~keV, similar to the initial peak, the emitting radius is around 3.3$\pm$0.4~km. This value corresponds to roughly 51\% of the radius measured during the initial peak.  
In the case of burst II, at the peak the temperature reaches to  2.44$\pm$0.13~keV, with an apparent emitting radius of 6.19$\pm$0.4~km. The peak flux of the secondary peak reaches to F$_{peak}$=1.4$\pm$0.2$\times$10$^{-8}$~\fluxcgs. The temperature and apparent emitting radius values at the secondary peak are 1.66$\pm$0.13~keV and 5.3$_{-0.6}^{+0.5}$~km, respectively. These values indicate that the secondary peak, in this case, reaches to a  somewhat lower temperature, while covering a similar area on the surface compared to its initial peak. On the other hand, for burst III peak flux is found to be F$_{peak}$=12.64$\pm$1.42$\times$10$^{-8}$~\fluxcgs and the peak temperature reached to 3.21$\pm$0.25~keV. These values are significantly higher than the previous bursts, reported here and indicate the possible affects of accretion rate on the properties of thermonuclear bursts.

\begin{figure}
    \centering
    \includegraphics[scale=0.35]{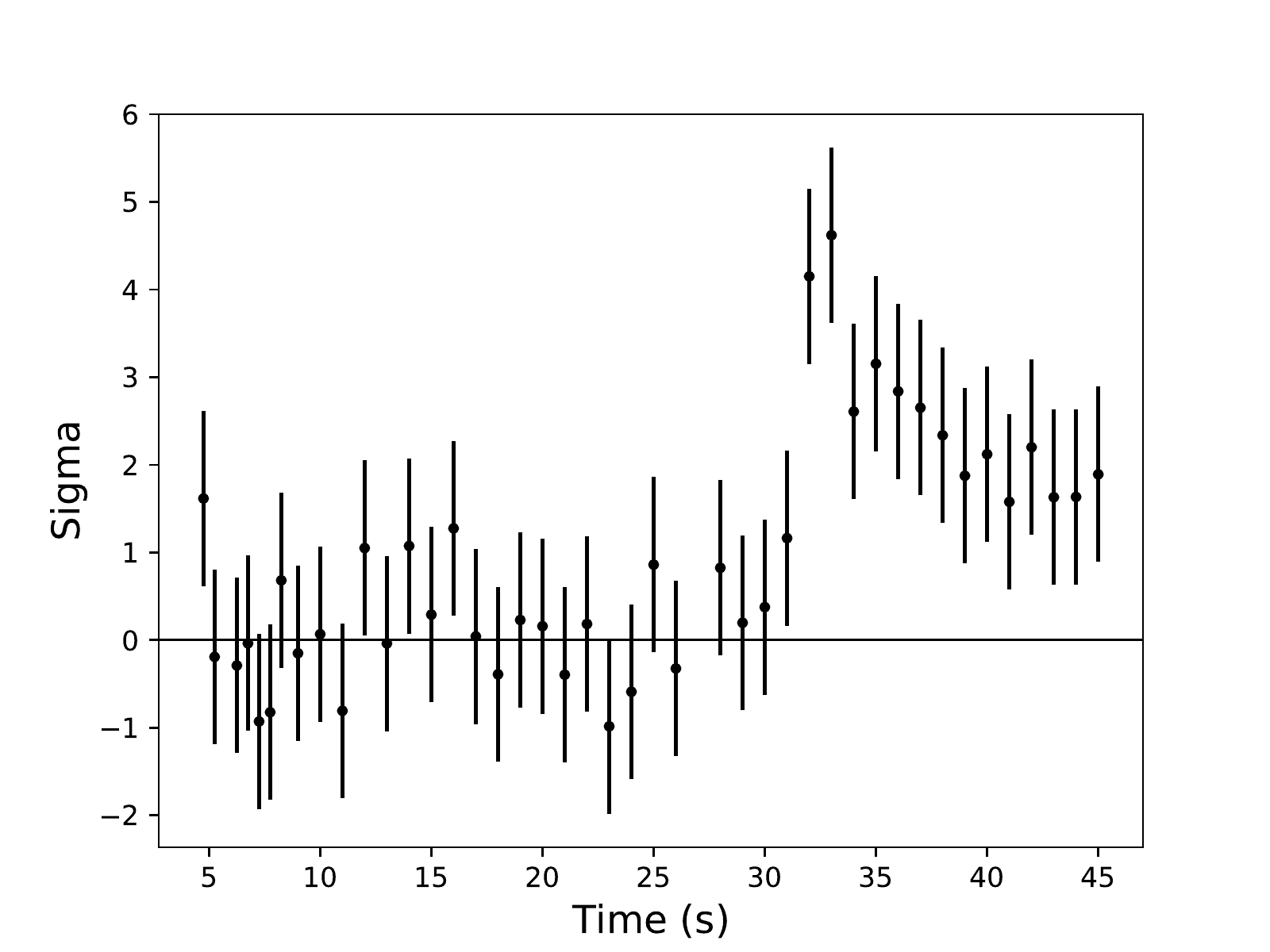}
    \includegraphics[scale=0.35]{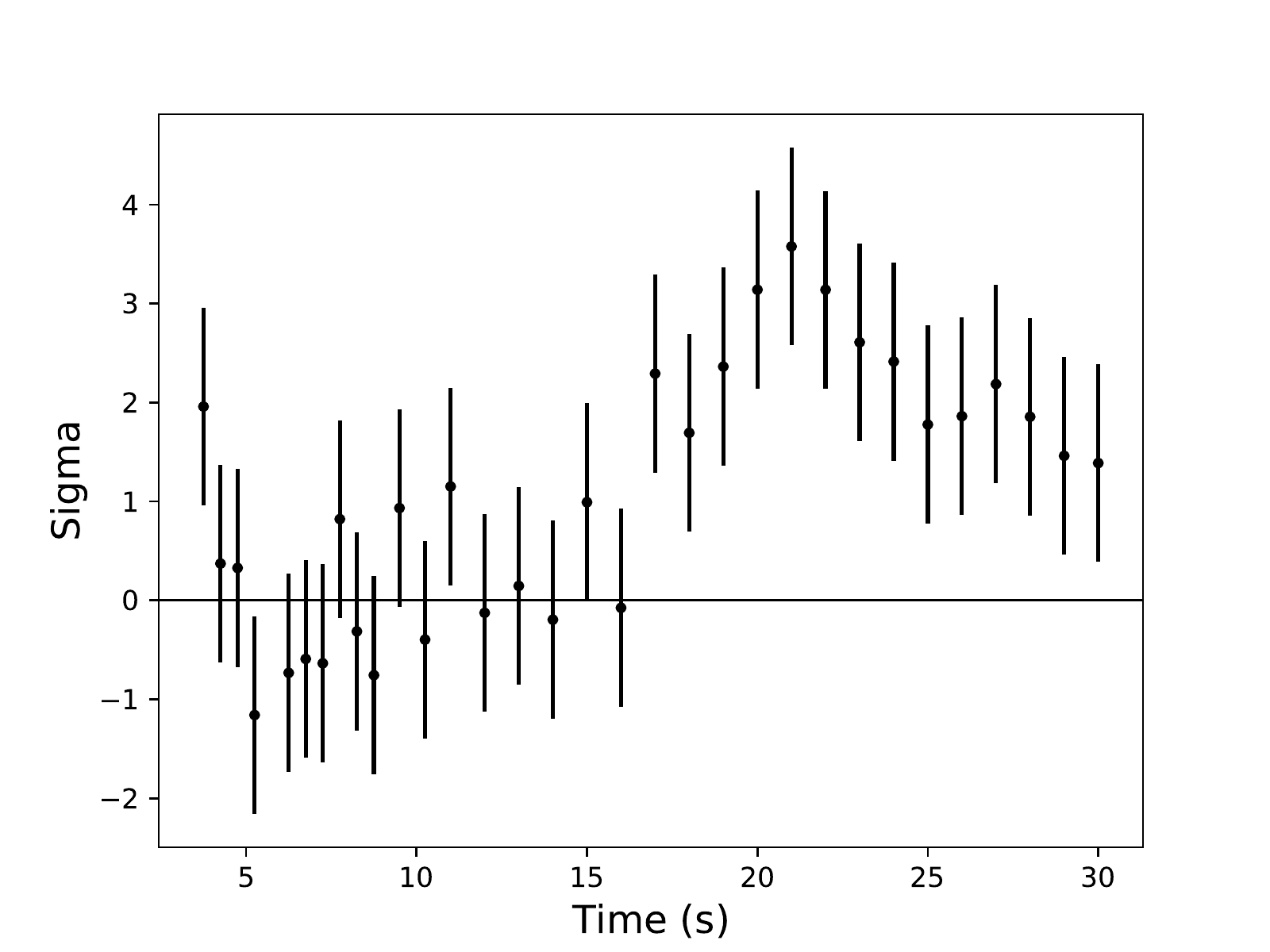}

    \caption{Residuals from an exponential evolution of the temperature in burst I (shown in the upper panel) and burst II. The heating and cooling during the secondary peaks is evident.}
    \label{fig:burst_kt_ev}

\end{figure}

If we fit the evolution of the temperature during burst I with an exponential decay from the moment the temperature started to decrease and just before the start of the secondary peak (4.75 s - 25 s), we find an e-folding time scale of 32$\pm$4~s. Following a similar method for burst II we found the temperature decay timescale of 27.8$\pm$5.5~s,  when we fit the data obtained within 3.8-15 seconds after the burst starts. The residuals from these fits are shown in Figure \ref{fig:burst_kt_ev} and indicate that in both of the secondary peaks we see a significant deviation from the expected decay. These results reveal that the secondary peaks we report here have their own spectral signature, which has not been observed before from this source.

Defining the burst duration as the time when the flux is greater than 5\% of the peak (excluding the persistent emission), the fluence of the initial peak of burst I can be found as $E_b=48.4\pm1.09\times10^{-8}$~erg~cm$^{-2}$, while for the secondary peak the fluence is  $E_b=9.11\pm0.67\times10^{-8}$~erg~cm$^{-2}$. A similar calculation for burst II is more complicated given the fact that the secondary peak happens while the flux is still higher than 5\% of the peak. For burst III the fluence can be found as $E_b=1.12\pm0.02\times10^{-6}$~erg~cm$^{-2}$.

\subsection{Search for Burst Oscillations} \label{sec:burstoscillation}

We performed a deep dynamic search for burst oscillations throughout all X-ray burst events. For burst I, we selected all events in the 0.5$-$8.5 keV band from 90~s time segment, starting from 20~s prior to the peak time. We then performed a search for each 4-s long segment using the Z$^2$ statistic with one harmonic \citep{buccheri} within the frequency interval of 618 to 622 Hz. We slid 4-s long search window with a one-second increment and repeated the search in the same frequency range for each time segment. We finally constructed a dynamic power spectral density for the entire search interval. We find no episode in burst I with Z$^2$ power corresponding to the single-trial significance of at least $4 \sigma$, indicating that there are no burst oscillations. A similar search was performed for burst II in a 70~s long time interval, again starting from 20~s prior to the burst peak. The resulting power spectral density yields no episode in burst II with pulsations having single-trial significance more than $4 \sigma$, indicating an absence of burst oscillation for this event as well. Finally, using a similar method, we looked for burst oscillations in burst III as well starting from 20~s before the onset, up to 70~s after the start. Similarly we found no significant oscillation candidate in this event. 
The upper limits on the amplitude of burst oscillation during these three bursts were calculated following the procedure described in \citep{2017ApJ...834...21O}. In the case firstly we took the measured power as the maximum power obtained in the frequency range searched. Considering this measured power, we then estimated the signal power as the one for which the probability of observing the measured power given the signal power was found to be maximum. The upper limit was then computed using the relation
\begin{equation}
 A_{\rm rms} = \sqrt{\frac{Z_s^2}{N}} \left( \frac{N}{N-N_b} \right )
\end{equation}
where $ A_{\rm rms}$ is the upper limit on the rms fractional oscillation amplitude, $Z_s^2$ is the signal power, $N$ is the total counts and $N_b$ is the background counts. Here the upper limit is calculated in a similar method to that for the detections and thus can be readily compared to the detected burst oscillation amplitudes for this source in the previous literature. The upper limits for the events were found to be 3.40 \%, 2.78 \% and 6.21 \% respectively. These upper limits are consistent with the upper limits previously observed from this source when no oscillation was detected and is typically less than the oscillation amplitudes observed in the case of significant detection \citep[see e.g., Figure 4 of][]{2017ApJ...834...21O}.

\section{Discussion and Conclusions} \label{sec:dicussion}

In this paper, we report the detection of two thermonuclear X-ray burst events in June 2020, while \src was at the peak of an outburst and a third one which happened just at the end of the same outburst in September 2020. In the first two bursts, we clearly detect secondary burst-like events, 30~s and 18~s after the initial peaks. Analysis of the light curves implies that these secondary peaks also show a fast rise exponential decay trend. Furthermore it is possible to see the traces of the secondary peaks in the results of the time resolved spectral analysis.

Unlike the secondary peak observed in burst I, the secondary peak observed in burst II shows a more gradual change in the spectral parameters (see Figures \ref{fig:bursts_method1} and \ref{fig:burst_kt_ev}). The lack of a sharp transition in the temperature evolution can also be manifested due to the low flux exhibited by the secondary bursts. A similar phenomenon was shown by \citet{2011ApJ...733L..17L} where for bursts having low values of peak luminosity to persistent luminosity ratio it was observed that the spectral evolution shows a more gradual variation. Such bursts with gradual temperature and normalization evolution along with smaller effective radii are typically observed at high accretion rates which is consistent with the high accretion rate obtained for these observations.

Since its discovery, \src has been one of the most active and bright transient low mass X-ray binary and provided a wealth of data. Despite the fact that more than 150 X-ray bursts have been reported from this source, bursts with recurrence times less than a few minutes have neither been reported from this source nor even from other X-ray bursters (the shortest recurrence time reported has been around 3-6 minutes \citep{2010ApJ...718..292K,2012ApJ...748...82L}).

\begin{figure*}
    \centering
   \includegraphics[scale=0.35]{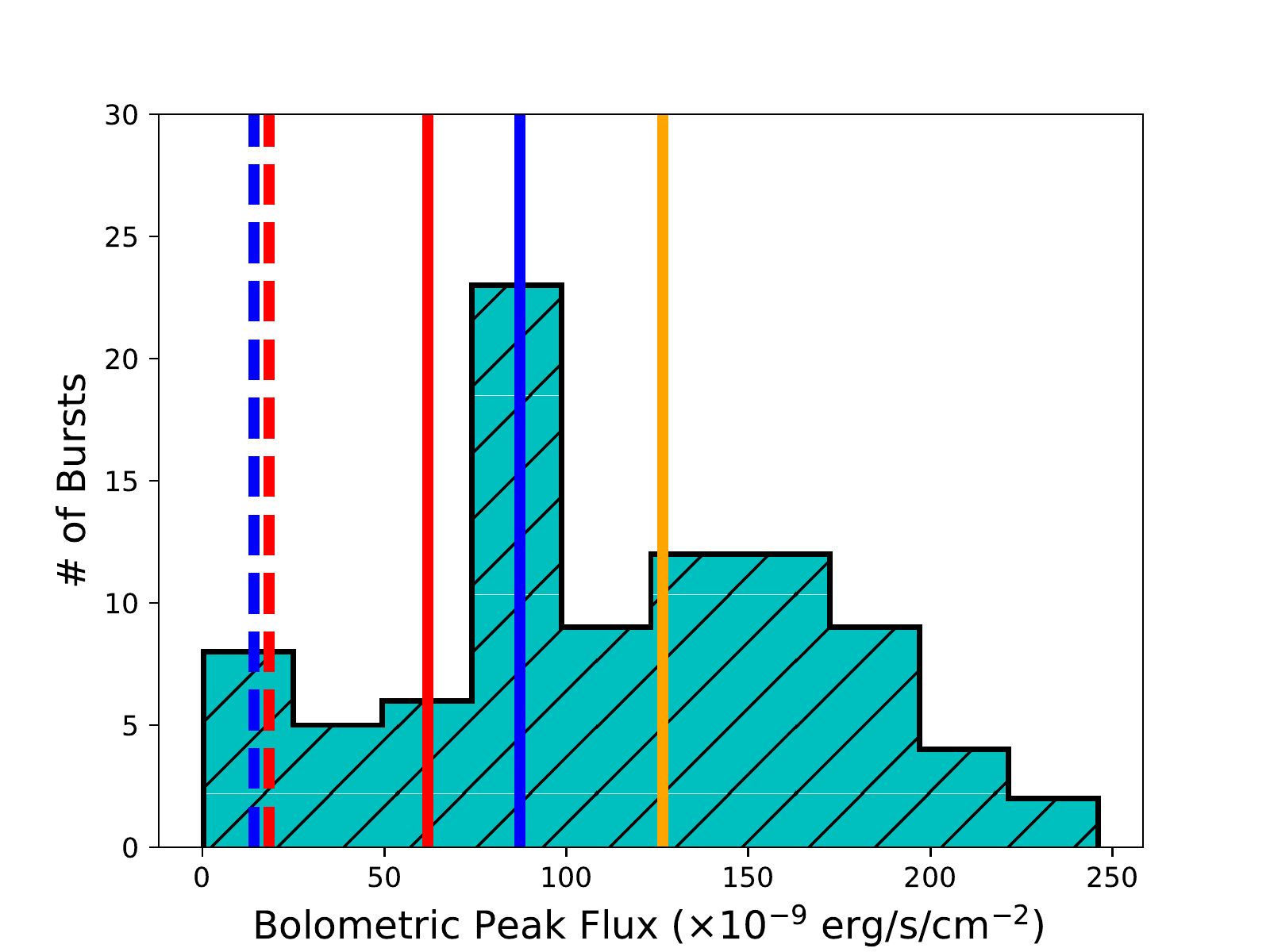}
    \includegraphics[scale=0.35]{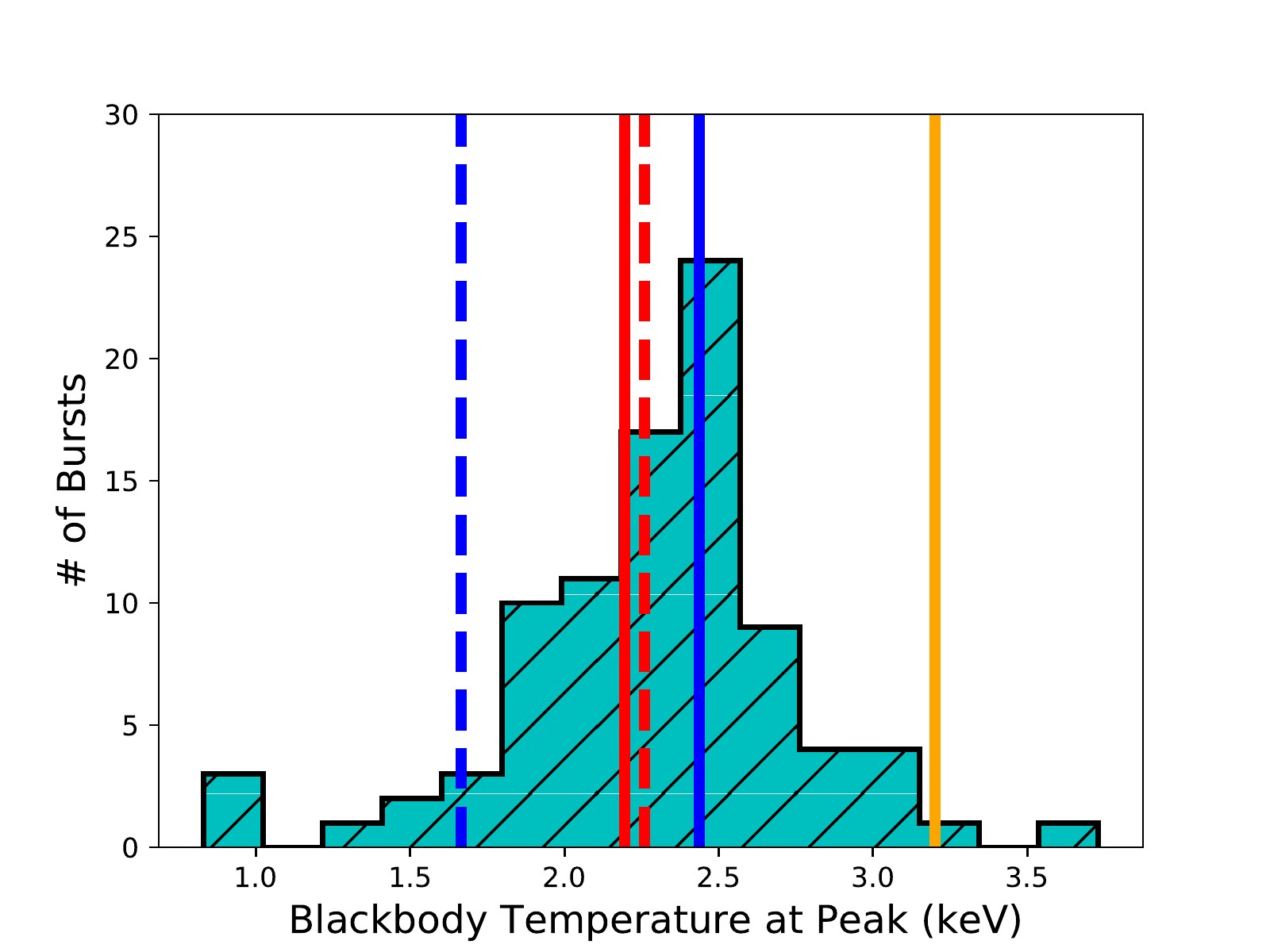}
    \includegraphics[scale=0.35]{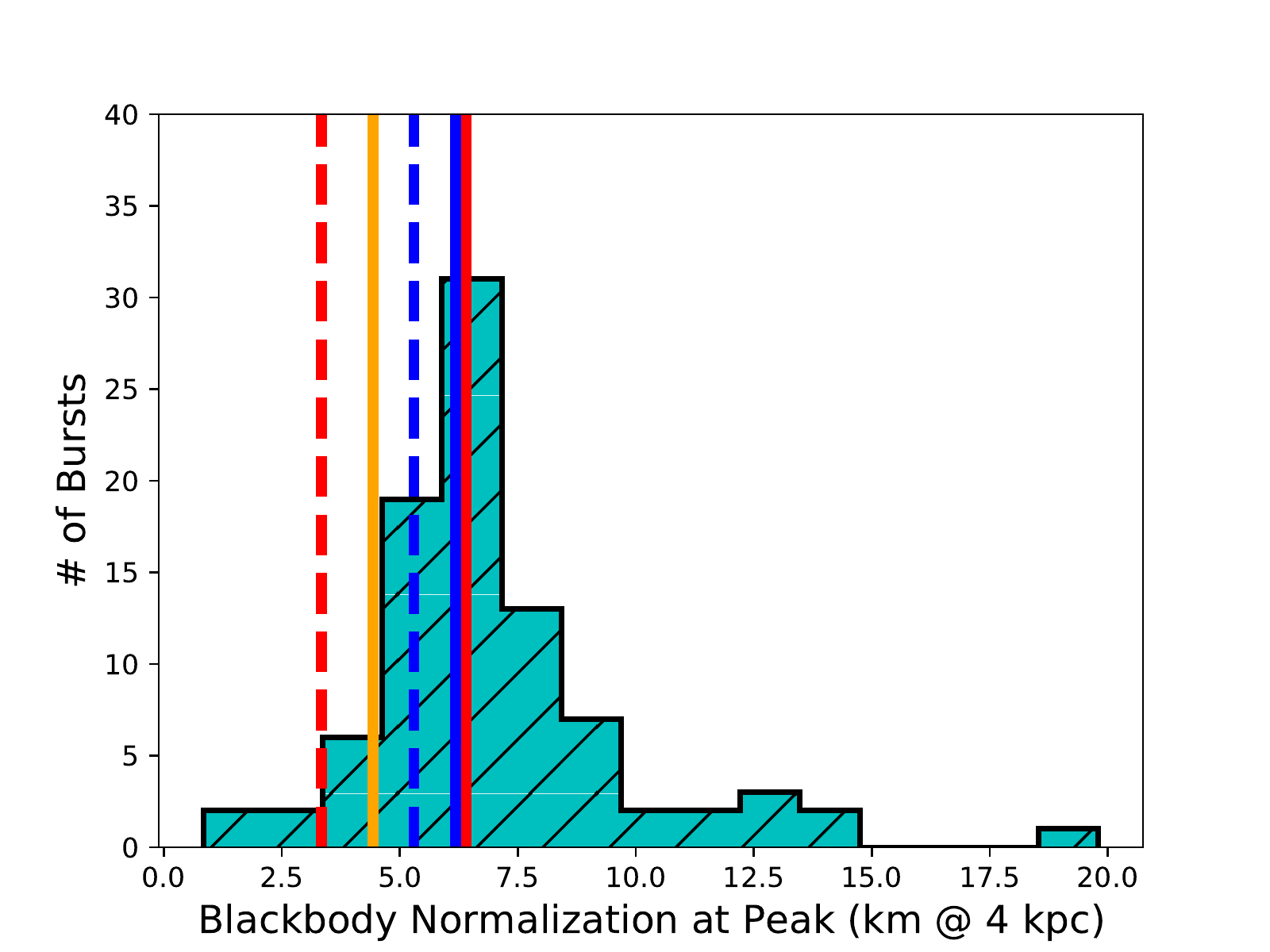}

    \caption{From left to right histograms of observed  peak bolometric flux, blackbody temperature and blackbody normalization values at the peak flux for \src, as in MINBAR catalog, respectively. Vertical blue, red, and orange  lines show the values measured 
    for burst I, II, and III, respectively. The same values for the secondary peaks are shown with vertical dashed lines of the same color, respectively.}
    \label{fig:bursts_hist}

\end{figure*}

To better assess the characteristics of the X-ray bursts reported here, and to put the secondary peaks in perspective, we compared them with the ones cataloged in the MINBAR catalog \citep{2020ApJS..249...32G}. In Figure \ref{fig:bursts_hist} we show the histograms of peak flux, blackbody temperature, and normalization at the peak flux for all the X-ray bursts recorded in the MINBAR catalog. The distribution of peak flux, blackbody temperature, and blackbody normalization at the peak flux values for all the  X-ray bursts observed from \src are shown in Figure \ref{fig:bursts_hist}. The values we obtain from spectral fits are also shown and indicate that while the initial peaks are in general very much inline with the rest of the X-ray bursts observed previously from this source, the values for the secondary peaks are located towards the lower end of the previously observed values. Especially, the temperature for the secondary peak of burst II and the blackbody normalization value obtained for the secondary peak burst I is at the lowest end of the observed values. Note that best fit parameters for burst III are more in line with previous bursts with the exception of the peak temperature value, which seems to be one of the highest. However, soft X-ray response of \nicer~ does not allow for a precise determination of such a high temperature, therefore we have a relatively large (8\%) uncertainty in this parameter.

First two X-ray bursts we report here are observed at the peak of an outburst from \src. We calculate that during this state the persistent flux of the source was roughly 15\% of the Eddington luminosity for the source.  For burst III the persistent flux was only 0.9\% of the Eddington limit. Within the MINBAR catalog there are only three X-ray bursts (MINBAR burst IDs, 1599, 3015, 8002) out of the total of 147, observed from \src when the $\Gamma$ value was above 14\%,  where $\Gamma$ is defined in the catalog as the ratio between the persistent flux of the source and the peak photospheric radius expansion burst flux. The small number of bursts seen at such high accretion rate from \src shows that the detection of these bursts is already exceptional. Within the small sample we were able to identify in the MINBAR catalog, no short recurrence or even a short waiting time burst event is reported.

On the other hand, within the MINBAR catalog five X-ray bursts observed from \src have recurrence times less than 0.1~h. Within these X-ray  bursts in cases where it could be determined, the $\Gamma$ value is reported to be 1-3\%. Such low $\Gamma$ values indicate that SWT bursts observed from \src happen at much lower accretion rates. We note that \cite{2014ApJ...787..101K} show that, in fact all SWT bursts occur in hard/intermediate state, while the bursts we report here occur during a soft state {(see Figure \ref{fig:hid_ccd})}. Although this does not necessarily argue against the secondary peaks here being SWT bursts, it does show that these events should be considerably different given the fact that at the soft state the accretion geometry is expected to be different compared to the hard states.

Secondary peaks in X-ray bursts from \src has been reported before  \citep[see, e.g.,][]{1989A&A...208..146P, 2014MNRAS.442.3777P, Jaisawal2019}. Specifically, bursts 10, 15, and 17 in \cite{2014MNRAS.442.3777P}, the X-ray burst observed in 1986 and reported by \cite{1989A&A...208..146P} or the X-ray burst reported by \cite{Jaisawal2019}, do have secondary peaks in their lightcurves. However, the  X-ray bursts we report here show significant variations compared to those events. First, the secondary peaks reported earlier all seem to occur very close to the initial peak of the burst. Second, in all these previous cases the accretion rate was much lower compared to the events we report here. Burst III we report here happened at such low accretion rates but does not show any evidence for a secondary peak. Last but not least, in all the previous cases there is no notable spectral variation reported apart from the typical spectral evolution already observed during the bursts that do not show a secondary peak. In the cases we report here, the secondary peaks are significantly late, show their own spectral evolution (i.e., heating and cooling as shown in Figure \ref{fig:burst_kt_ev}) and the X-ray bursts happen at much higher accretion rates.

One of the scenarios that can explain the secondary peak after burst I may be based on partial burning on the surface since it reaches to a temperature that is very close to the value obtained for the initial peak but covers a smaller apparent emitting radius. Possible mechanisms for X-ray bursts covering only smaller areas on the surface have been proposed and mostly rely on the effect of magnetic fields on the accreted material or fast rotation \citep{1988A&A...198..163F,2009ApJ...706..417L}. Short recurrence time bursts have been shown to be observed mostly from fast-spinning neutron stars \citep{2010ApJ...718..292K}. Observed  X-ray burst oscillations at around 620~Hz \citep{2002ApJ...580.1048M} indicate that \src hosts one of the fastest-spinning neutron stars in a low mass X-ray binary.

 It is generally assumed that during an X-ray burst, the burning front can propagate to at least a large fraction of the surface within a few seconds time scale \citep{2002ApJ...566.1018S}. However, magnetic field and stellar spin are also expected to play a significant role on the characteristics of an X-ray burst. During some of the X-ray bursts, burst oscillations have been observed from a number of LMXBs \citep[for a review see,][]{2012ARA&A..50..609W}. While the observations of oscillations during the rises of X-ray bursts are generally attributed to the spreading of the burning front, the physical mechanism behind the oscillations observed during the decay phases of the bursts are not so clear. Models employ the magnetic field acting on the surface \citep{2006ApJ...641..471P} or the effect of Coriolis forces for the rapidly rotating neutron stars \citep{2006ApJ...636L.121B, 2007ApJ...666L..85B} or the global surface modes \citep{2012ARA&A..50..609W} to stall or limit the burning to a confined region on the surface. \cite{2006ApJ...641L..53B} suggests that X-ray bursts starting at or near one of the poles may slow or even stall as the burning front approaches the equator, before speeding up again as it propagates to the opposite pole. The mechanism for the stall is suggested to be the freshly accreted material towards the equator \citep[see also,][] {2015MNRAS.448..445C}. Although an upper limit on the separation of the peaks is not given by \cite{2006ApJ...641L..53B} it is expected to be of the order of several seconds. The secondary peaks we report here have separations that are significantly more than what is expected from a stalling mechanism on the surface of the neutron star. Furthermore, models for double-peaked X-ray bursts often predict that the intensity in the secondary peaks would be comparable to the initial peaks. Again, this is not the case in the bursts we report here, since the secondary peaks only reach roughly two thirds or even lower of the initial peaks. Finally, a triple-peaked X-ray burst has been observed from \src \citep{2009MNRAS.398..368Z}, which may be hard to explain with the flame stalling scenario discussed here. On the other hand, we note that recently \citet{2017ApJ...851....1C,2020MNRAS.tmp.2718C} showed that in these systems the ignition latitude shifts to the poles of a neutron star as a function of mass accretion rate. The fact that the bursts we report here occur at high accretion rates, supports the near-polar origin. 
 
Another possible scenario behind secondary peaks can be the modified ignition conditions set by the preceding burst. The bright initial peak may result in residual effects on the envelope long after the burst peak and at this high accretion rate, this may lead to a trigger of unstable ignition following a much shorter accumulation of fuel. Such successive burst triggering with less accumulated fuel has been proposed by \citet{1992ApJ...390..634F}, \citet{1993ApJ...413..324T}, and \citet{1996ApJ...459..271T}, especially for X-ray bursts with long tails. The residual heating of the envelope from the preceding  X-ray burst ensures that a smaller ignition column depth is required to initiate the thermonuclear ignition process. For the case of the two burst events presented here, it may be that the accretion rate is high enough that the threshold ignition column depth is being achieved even during the decay tails of the initial peaks, perhaps giving rise to the secondary peaks. The exact ignition conditions will be critical in generating such short recurrence bursts. 

To summarize, we observed three typical thermonuclear X-ray bursts from \src during the outburst it exhibited in 2020 with NICER. However, in two of these X-ray bursts, we clearly detected secondary peaks 30~and~18~s after the initial peaks. In both cases, we see spectral variation especially an increase in the temperature followed by cooling in addition to a change in burst flux. The time difference between the initial and the secondary peaks are too short to be explained by scenarios employed usually to explain SWT bursts, on the other hand, they are also too long to be easily understood in terms of flame propagation scenarios. Furthermore, the peak fluxes in the secondary peaks are significantly low compared to the initial peaks, which is expected in the double-peaked scenarios. We conclude that the  X-ray bursts reported here challenge our current understanding of thermonuclear X-ray bursts and show the necessity to perform realistic models for ignition of the bursts and the propagation of the burning front.

\section*{ACKNOWLEDGMENTS}
We thank the referee very much, whose comments and suggestions improved the manuscript. T.G. has been supported in part by the Scientific and Technological Research Council (T\"UB\.ITAK) 119F082, Royal Society Newton Advanced Fellowship, NAF$\backslash$R2$\backslash$180592, and the Turkish Republic, Presidency of Strategy and Budget project, 2016K121370. C.M. is supported by an appointment to the NASA Postdoctoral Program at the Marshall Space Flight Center, administered by Universities Space Research Association under contract with NASA. DA acknowledges support from the Royal Society. This work was supported by NASA through the NICER mission and the Astrophysics Explorers Program.
This research has made use of the MAXI data provided by RIKEN, JAXA and the MAXI team as well as the data and software provided by the High Energy Astrophysics Science Archive Research Center (HEASARC), which is a service of the Astrophysics Science Division at NASA/GSFC and the High Energy Astrophysics Division of the Smithsonian Astrophysical Observatory.

\facilities{NICER}
\software{HEASoft, XSPEC, Sherpa, CIAO, Astropy}

\bibliography{sample63}{}
\bibliographystyle{aasjournal} 

\end{document}